# A Simple Minkowskian Time-Travel Spacetime

John D. Norton

Department of History and Philosophy of Science

University of Pittsburgh

## Abstract

This relativistic, time-travel spacetime is everywhere metrically flat, excepting a conical singularity. Observers following timelike geodesics can eventually encounter their past selves, aging in the opposite time sense. The spacetime is not time-orientable.

## 1. Introduction

Time travel fiction commonly depicts time travelers who encounter their past selves or, in the grandfather paradox, their ancestors. In traditional fictional representations of time travel, such as in H. G. Wells's, *The Time Machine*, travelers age in the same time sense as those visited in the past and future.[1] Elsewhere, fantasy fiction supplies another possibility, The wizard Merlyn in T. H. White's 1938 fantasy novel, *The Sword in the Stone*, meets a young Arthur. Merlyn ages in the opposite time sense to Arthur.[2] Arthur's first meeting with Merlyn is Merlyn's last meeting with Arthur; and Arthur's last meeting with him is Merlyn's first. We can imagine time travelers who arrive in the past to meet their former selves, but now age in the opposite time sense. They are still time travelers since they are meeting their past selves. But we have now added a twist from another part of the fantasy literature.

We may doubt whether such differences in aging conforms with what we know of thermodynamics, statistical physics and even biology. Such was Einstein's reaction to Gödel's original description of his time-travel universe.[3] [4] The possibility of signaling from the future to the past, Einstein noted, contradicts the inexorable rise in entropy for real processes in thermodynamics. It has been traditional since Gödel's paper to ask the simpler question of whether general relativity admits spacetime structures that, independently of further physics, allow time travel. We can ask the corresponding question of whether general relativity admits spacetime structures in which an aging Merlyn might meet his younger self, while they age in different time senses. The affirmative answer is supplied in this paper.

Einstein's general theory of relativity does admit spacetimes in which time travel is possible, in the sense that they harbor closed timelike curves. That this is so has been known since at least Gödel's time travel solution of Einstein's $\lambda$-augmented gravitational field equations. Since then, many other time-travel spacetimes have been found within Einstein's theory, such as are afforded by Kerr black holes. They have attracted considerable attention in both physics and philosophy.[5] Some proposals require exotic physics[6], to open a wormhole that connects different parts of spacetime. Others escape this complication in wormhole creation by simply stipulating a topology altering connection between two parts of the spacetime.

This paper presents one of the simplest time-travel universes admitted by Einstein's general theory of relativity. It is matter free and everywhere metrically flat, except for a singular, two-dimensional surface around which timelike geodesics are deflected back towards their past. As a preview, Figure 1 is a caricature of the time travel spacetime. The spacetime to the left of the figure is roughly Minkowskian and is more accurately so as we go farther left. The usual time direction there is left-right, as indicated by the disposition of the light cones. The spaceships shown are moving inertially along timelike geodesics that lie within these light cones. No rocket motors fire. As they approach the singularity on the right of the figure, their geodesics are deflected so that their motion in time is reversed, whereupon they return to the spacetime on the left of the figure. There, time travelers encounter their past selves who age in the opposite local time sense. Merlyn, traveling in such a spaceship, would be able to communicate with his past self, by sending him a light signal as shown.

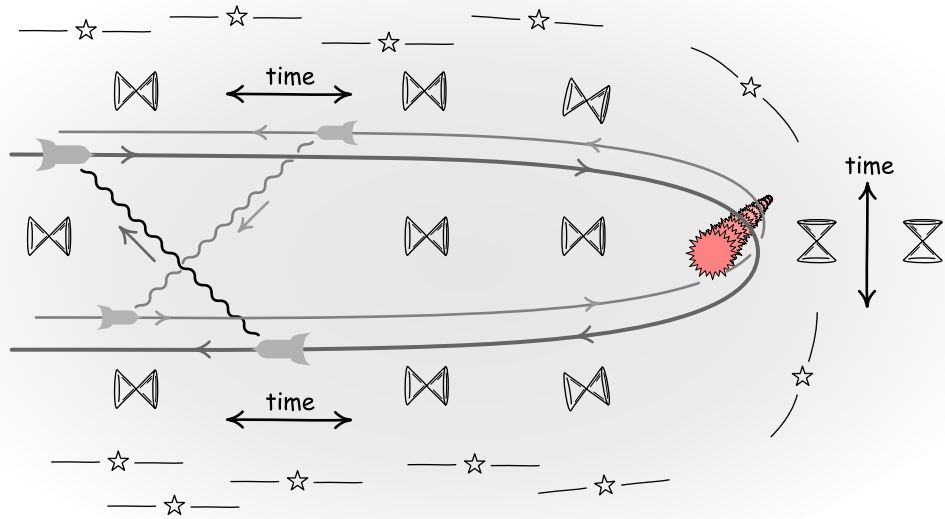

Fig. 1. Time travel in the time-travel spacetime

If the spaceship accelerates when it nears its past self, its timelike curve would close in the sense that it intersects with itself. The resulting closed curve is not a "closed timelike curve" of the traditional time travel literature in general relativity, since the curve in this new spacetime must have at least one event at which the curve is not differentiable and thus has no tangent vector; or, fail that, one event at which the curve has become spacelike. This difference is a manifestation of the fact that the spacetime is not time orientable. That is, no consistent division of timelike motions into future and past is possible. Any such division will be contradicted by the return of the motion after it passes the singularity.

The example is offered as a pedagogically useful addition to our repertoire of time-travel universes. In this pedagogical application, it has several interesting features.

It is a natural student question to ask how time travel comes about in such time-travel spacetimes. The simplest case takes a Minkowski spacetime and merely identifies two spacelike hypersurfaces, so that the present evolves back to itself. This "cylinder" universe has the global topology of $\mathbb{S} \times \mathbb{R}^3$. While such stipulations produce spacetimes that are, in my view, admissible within Einstein's theory, one could be forgiven for the sense that time travel has been introduced artificially into the theory by our meddlesome stipulation as opposed to a sound reason of physics. Earman et al. in Ref. 5 have reviewed the difficult and still open question of whether we could do something that might bring about a time machine, that is, bring about closed timelike curves. At least some time-travel universes are so structured that we can point to a cause of the temporal anomaly. In Tipler's proposal[7], frame dragging effects are produced by a rapidly rotating cylinder of matter and they are sufficient to lead to closed timelike curves.

In the spacetime of this paper, as Figure 1 shows, we can, if we are so inclined, attribute the possibility of time travel to the disturbing influence of the singularity.[*] The spacetime is otherwise unremarkable, in being everywhere locally flat, like the spacetime of special relativity.

Section 2 below gives a specification of the intrinsic geometry of the spacetime, including the intrinsic metric, its metrical curvature, the disposition of its timelike geodesics and its light cone structure. Providing this specification usually requires more elaborate computations, such as setting up and solving applicable differentiable equations. Sections 3 to 6, however, show that the entire specification can be recovered by elementary means involving little more than coordinate transformations. The pedagogic value of these constructions is that they give a more immediate sense of how the time travel properties of the spacetime come about.

Finally, the paper addresses a more sophisticated problem of pedagogic interest. A conical singularity is singular in its *extrinsic* curvature. What of the geometry *intrinsic* to the surface? In what sense is it singular? The spacetime geometry is metrically flat in all neighborhoods arbitrarily close to the apex. It follows that the intrinsic metric is flat all the way along any path leading to the singularity. Thus, it would seem that taking the limit along the paths to the singularity would assign a flat spacetime intrinsic metric to it. Section 7 shows in detail that the limits taken along all paths do not converge to a unique metric, which establishes the sense in which a singularity is present.

---

[*] Such attributions are, in my view, purely of heuristic value. That they seek a notion of causal influence over and above the relations already provided by the prevailing physical theory is yet another of the many attempts at a priori physics, all of which have met with little success over millennia. See John D. Norton "The Metaphysics of Causation: An Empiricist Critique," pp. 58-94 in Yafen Shen, ed., *Alternative Approaches to Causation.* (Oxford: Oxford University Press, 2024).

## 2. The Spacetime

The geometrical structure of the spacetime is given by the line element for the interval $s$:

$$ds^2 = (r^2/4) \cos\theta \, d\theta^2 - \cos\theta \, dr^2 + r \sin\theta \, d\theta \, dr - dy^2 - dz^2 \qquad (1)$$

where polar coordinates $r$, $\theta$ have values $r > 0$ and $0 \leq \theta < 2\pi$. Cartesian coordinates $y, z$ have values $-\infty < y, z < \infty$. The metrical coefficients in a coordinate basis $x^i = (\theta, r, y, z)$ are:

$$g_{ik} = \begin{bmatrix} \frac{r^2}{4}\cos\theta & \left(\frac{1}{2}\right) r \sin\theta & 0 & 0 \\ \left(\frac{1}{2}\right) r \sin\theta & -\cos\theta & 0 & 0 \\ 0 & 0 & -1 & 0 \\ 0 & 0 & 0 & -1 \end{bmatrix} \qquad (2)$$

This coordinate system, shown in Figure 2, "goes bad" at the origin $r = 0$, since an event there would be assigned all values of the angle coordinate $\theta$. It will turn out, as shown in Section 7 below, that there is a singularity at $r = 0$ in the sense that taking a continuous limit from the surrounding spacetime provides no unique extension of the metrical structure to events that we might suppose to be labeled by $r = 0$. For all events other than at $r = 0$, it will become apparent in Section 3 that the spacetime is metrically flat with the usual Lorentz signature of a Minkowski spacetime. That is, any open, $\mathbb{R}^4$ neighborhood of the spacetime excluding $r = 0$ is isometric with an open $\mathbb{R}^4$ neighborhood of a Minkowski spacetime. Thus, it satisfies Einstein's unaugmented, source-free field equations with vanishing curvature everywhere, except at $r = 0$. However, as Figure 1 already suggests, the spacetime is not time orientable.

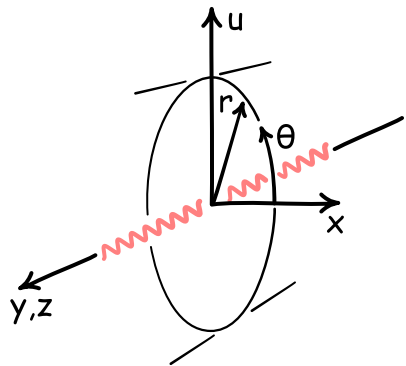

Fig. 2. Cylindrical coordinates of the spacetime and its singularity

In Figure 2, the singularity at $r = 0$ is drawn as a line within a three-dimensional space. The figure suppresses one dimension of the four-dimensional spacetime by collapsing the $y$- and $z$- axes. Thus, the singularity is really a two-dimensional surface enclosed within a four-dimensional space.

It follows immediately from Eqs. (1) and (2) that the metrical structure is translation invariant in the *y* and *z* directions. That is, $y \to (y + \text{constant})$ and $z \to (z + \text{constant})$ are both isometries. The interesting, time-travel related physics happens in surfaces of constant *y* and *z*; that is in surfaces spanned by the coordinates *r* and $\theta$.

We shall see in a simple construction given in Section 5 below that two families of intersecting lightlike curves in these surfaces are given by:

$$r = \frac{k}{\sqrt{1-\sin\theta}}, \quad r = \frac{k}{\sqrt{1+\sin\theta}} \tag{3}$$

for $k > 0$. The lightcones adapted to these curves are illustrated in Figure 3 for a surface spanned by the coordinates *r* and $\theta$.

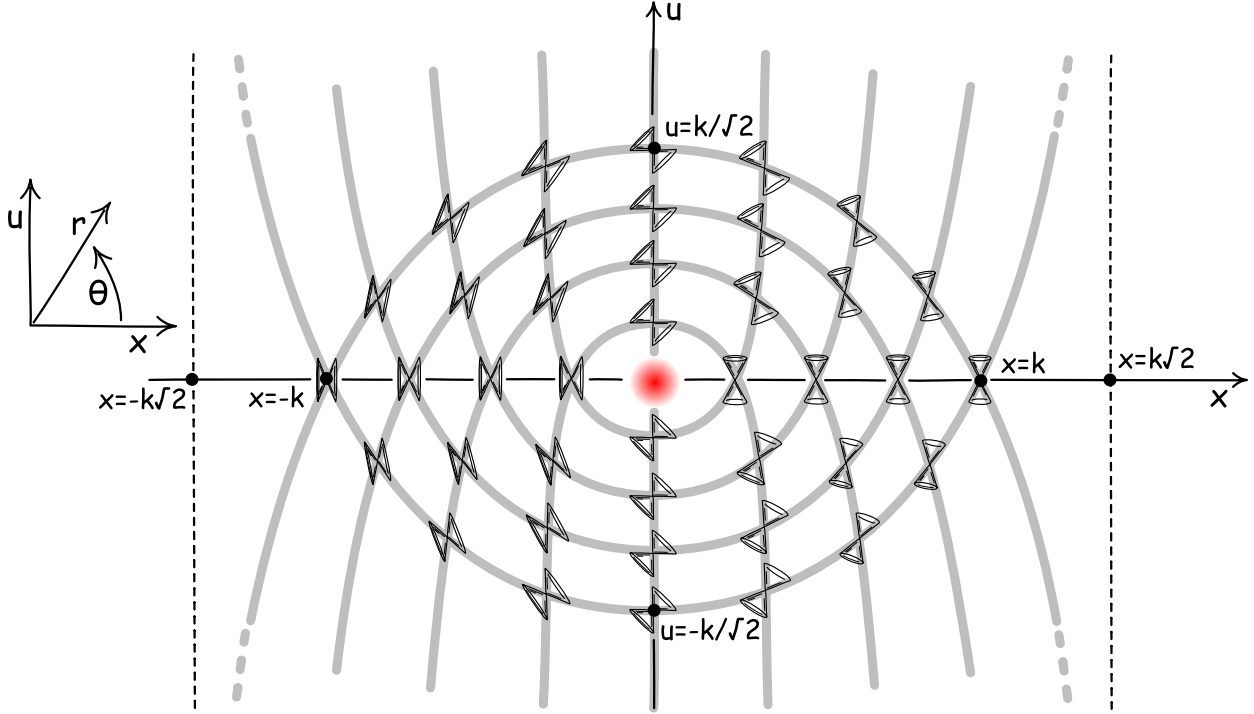

Fig. 3. Light cone structure in a surface of constant *y* and *z*

The Cartesian coordinates shown in Figures 2 and 3 are defined by

$$u = r \sin\theta \quad x = r \cos\theta \quad y = Y \quad z = Z \tag{4}$$

Briefly, at $\theta = 0$, the light cones indicate a timelike direction orthogonal to the radial direction. As we proceed in both the $+\theta$ and $-\theta$ directions, the light cones tip towards the singularity at $r = 0$. At $\theta = \pm\pi$, they meet such that timelike curves can pass directly into the singularity. If the lightlike curves intersect the *x* axis at $x = \pm k$, then they intersect the *u* axis at $u = \pm k/\sqrt{2}$ and are asymptotic to $x = \pm k\sqrt{2}$.

The region of spacetime for very negative *x* coordinates will appear much like an ordinary Minkowski spacetime. That is, for regions of very large, negative *x* in the vicinity of the *x*-axis, $\theta \approx \pi$, $\cos \theta \approx -1$ and $\sin \theta \approx 0$, the line element Eq. (1) then approximates the Minkowskian $ds^2 = dr^2 - dv^2 - dy^2 - dz^2$ if we introduce the new coordinate *v* such that $dv = r\, d\theta$. It is only for events near the singularity at $x = r = 0$ that the light cones tip toward the -*x*-axis and such that lightlike curves are eventually deflected around the singularity.

Timelike geodesics, similarly, behave much like in an ordinary Minkowski spacetime for the region with very negative *x* coordinates. As we enter regions close to the singularity at $x = u = r = 0$, the timelike geodesics are deflected around the singularity and reversed in their direction. More precisely, we shall see below that a family of timelike geodesics, for arbitrary fixed values of *y* and *z*, is given by

$$r = k/\cos(\theta/2) \tag{5}$$

where $k > 0$ is an arbitrary constant. The disposition of these geodesics is shown in Figure 4.†

---

† Caution is advised in reading these diagrams. The polar coordinates *r* and $\theta$ and Cartesian coordinates *u* and *x* do not have their usual metrical significance. Metrical judgments using them should be mediated by the line element Eq. (1) and metric components in Eq. (2).

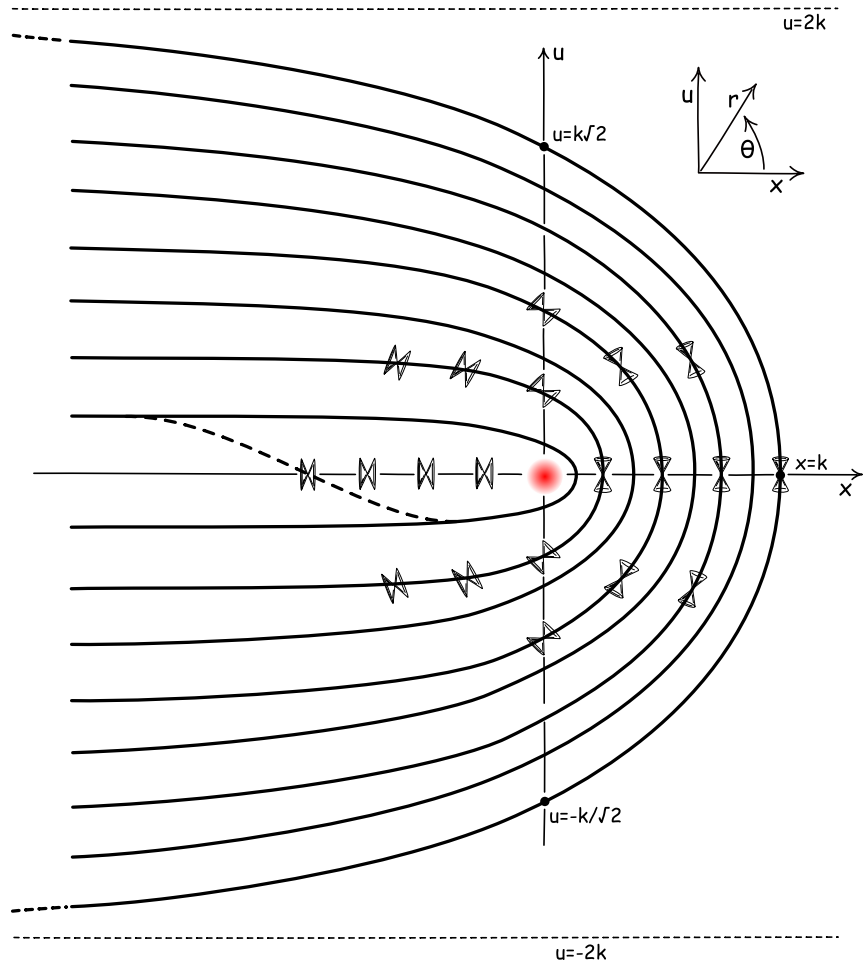

Fig. 4. A family of timelike geodesics

The geodesics intersect the *x* axis at *x* = *k*, the *u* axis at $u = k\sqrt{2}$ and, as they extend in the *-x* direction, are asymptotic to *u* = ±2*k*. While these geodesics do not form closed curves, if we adopt a position far enough in the *-x* direction, the two parts of each geodesic can be connected by another timelike curve, shown as a dashed curve in Figure 4. If we connect parts of these curves in obvious ways, we get closure in the sense of a single timelike curve that intersects its past self.

     An observer whose worldline coincides with one of these timelike geodesics, needs no acceleration to travel back in time. After sufficient proper time has passed, that observer will encounter the observer's past self, but aging in the opposite time sense.

     More generally, this spacetime is not time orientable in the usual sense of the existence of an everywhere non-vanishing, continuous, timelike vector field. That is, if we stipulate that some timelike vector, at some event, points in the future direction, parallel transporting that vector

along one of these timelike geodesics in both directions will eventually return it to a neighborhood where it has a contradicting time sense.

## 3 Constructing the Spacetime

The narrative so far has given no clue to the mode of construction of the spacetime. That has been done with the hope that the resulting spacetime will be assessed on its merits and not discounted because of the simplicity of the construction method. We arrive at the spacetime of Eq. (1) by an identification on a familiar Minkowski spacetime in the following way. This "source" Minkowski spacetime has the line element

$$ds^2 = dT^2 - dX^2 - dY^2 - dZ^2 \qquad (6)$$

where as usual $-\infty < T, X, Y, Z < \infty$. The "target" spacetime of Eq. (1) is recovered by introducing the coordinate systems of (1) and (4) in such a way as to cover the half of the source Minkowski spacetime specified by $X \geq 0$, as illustrated in Figure 5.

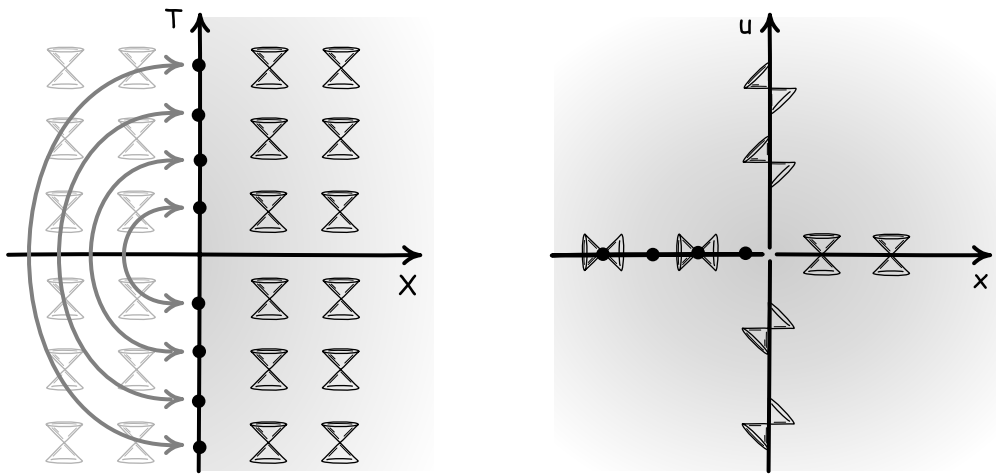

Figure 5. Construction of the time-travel spacetime

The construction requires identification of each event of Eq. (6) ($T$, $X$=0, $Y$, $Z$) with ($-T$, $X$=0, $Y$, $Z$). That is, the left half of the Minkowski spacetime, corresponding to $X<0$, is excised, as indicated on the left of Figure 5. Events connected by the double-headed arrows are identified to yield the target spacetime shown on the right of Figure 5.

To give further details, it is convenient to replace the $T$, $X$ coordinates of Eq. (6) with polar coordinates $r$, $\phi$, defined by

$$T = r \sin \phi \quad X = r \cos \phi \tag{7}$$

where $r > 0$ and $0 \leq \phi < 2\pi$. The mapping from the half-plane of the source Eq. (6) to the full plane of the target Eq. (1) is carried out by taking the metrical structure of the half Minkowski spacetime of Eq. (6) at the event $(\phi, r, Y, Z)$ and mapping it to the event $(\theta = 2\phi, r, y, z)$ in the target spacetime. Loosely speaking, the new metrical structure Eq. (1) is recovered by a doubling expansion of the angle variable about $\phi = 0$ of the source Minkowski spacetime. It can be written compactly as

$$\theta = 2\phi \tag{8}$$

## 4. Recovering the Metrical Structure and its Flatness

The line element Eq. (1) can be recovered from the line element Eq. (6) of the source Minkowski spacetime by a two-step transformation. First, cylindrical coordinates are introduced into Eq. (6) by the transformation Eq. (7). From Eq. (7) we have

$$dT = r \cos \phi \, d\phi + \sin \phi \, dr \tag{9}$$
$$dX = -r \sin \phi \, d\phi + \cos \phi \, dr$$

After substitution and some manipulation, the Minkowksi line element Eq. (6) becomes

$$ds^2 = r^2 \cos(2\phi) \, d\phi^2 - \cos(2\phi) \, dr^2 + 2r \sin(2\phi) \, d\phi \, dr - dY^2 - dZ^2 \tag{10}$$

The second step maps half the source Minkowski spacetime to the target spacetime by the substitutions:

$$\theta = 2\phi \quad r = r \quad y = Y \quad z = Z \tag{11}$$

The expression for the line element Eq. (10) becomes the corresponding expression Eq. (1):

$$ds^2 = (r^2/4) \cos \theta \, d\theta^2 - \cos \theta \, dr^2 + r \sin \theta \, d\theta \, dr - dy^2 - dz^2 \tag{1}$$

Since the source Minkowski metric is everywhere flat and the target spacetime is produced by a coordinate transformation, it follows that the new spacetime is also everywhere flat, excluding the singularity at $r = 0$.

## 5. Recovering the Light Cone Structure

This mapping Eq. (8) shows how deflections of the light cone structure of Figure 2 arise. We simply need to track how the light cones of the source half Minkowski spacetime are relocated and reoriented in the target spacetime under the mapping Eq. (8), as shown in Figure 6:

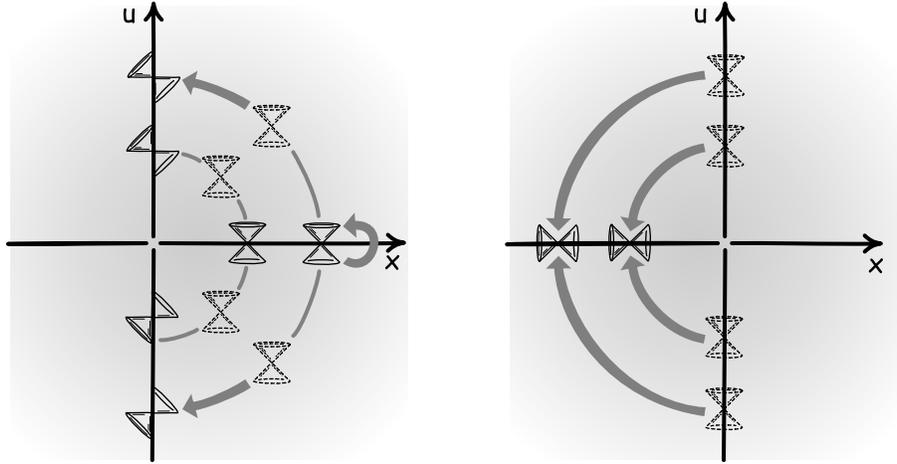

Figure 6 Mapping of light cone structure

As shown on the left of Figure 6, light cones on the positive *x*-axis of the spacetime are unchanged from those in the source spacetime along its positive *X*-axis. Light cones on the *u*-axis of the target spacetime are mapped from rotated light cones in the *X*>0 region of the source spacetime. As shown on the right of Figure 6, light cones on the negative *x*-axis of the target spacetime are mapped from rotated light cones on the *U*-axis of the source spacetime.

The mappings shown on the right of Figure 6 are two-to-one mappings. They concern the light cones mapped under $\phi = \pi/2 \rightarrow \theta = \pi$ and those mapped under $\phi = -\pi/2 \rightarrow \theta = -\pi$. Since $\theta = \pi$ and $\theta = -\pi$ are the coordinates of the same event (if the other coordinates are equal), it is essential that the two different mappings yield the same light cone. The two mappings deliver light cones such that one is the temporal inverse of the other. But since the light cones are time inversion invariant, the two mappings yield the same result. Hence the ensuing metrical structure is fully regular at all events $\theta = \pm\pi$ (where $r>0$). It is this inversion, however, that precludes the new spacetime being time orientable.

The analytic expressions Eq. (3) for the lightlike curves of Figure 2 are recovered from this mapping. First consider "future" (= +*T*) directed lightlike curves in the source Minkowski spacetime:

$$T = X + k \tag{12}$$

for constant $-\infty < k < \infty$. In polar coordinates introduced by Eq. (7), the curves are

$$r \sin \phi = r \sin \phi + k \tag{13}$$

or

$$r = \frac{k}{\sin\phi - \cos\phi} \tag{14}$$

where the range of values of $\phi$ to which this formula applies must be restricted to ensure that $r$ remains positive. (The case of $k = 0$ is excepted and addressed below.) Under the mapping Eq. (8) and similar angle restrictions, the expression becomes the first formula of Eq. (3)‡

$$r = \frac{k}{\sin(\frac{\theta}{2}) - \cos(\frac{\theta}{2})} = \frac{|k|}{\sqrt{1-\sin\theta}} \tag{15}$$

where the restriction to the absolute value of $k$ is all that is needed in the final $\sin\theta$ formula to ensure positive values for $r$. Applying this formula is complicated by the fact that *two* lightlike curves of the source Minkowski spacetime are mapped to form a *single* lightlike curve in the target spacetime. That is, the lightlike curves $T = X + k$ and $T = X - k$ for $k > 0$ are mapped under Eq. (8) to give a single curve of Eq. (3). These curves are illustrated in Figure 7.

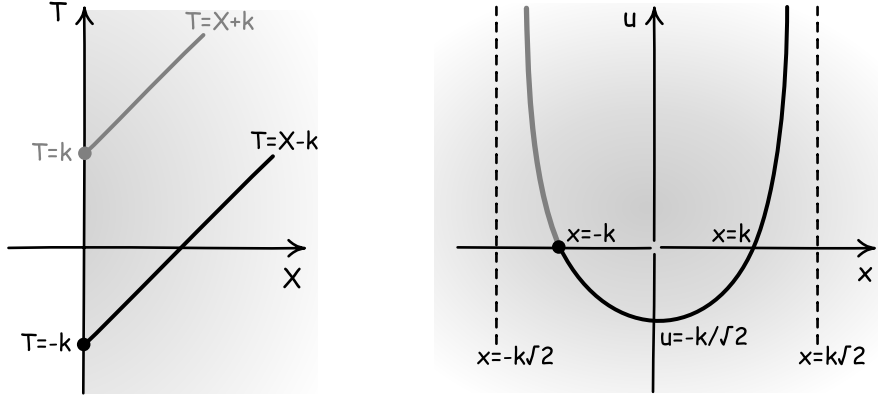

Fig. 7. Mapping of lightlike curves

The mapping of the two lightlike curves from the source Minkowski spacetime joins at $u = 0$, $x = -k$ in the target spacetime to yield a single curve. The resulting lightlike curve in the target spacetime diverges in the $+u$ direction. That it is asymptotic to $x = \pm k\sqrt{2}$ cannot be recovered directly from Eq. (3). It can be affirmed by re-expressing Eq. (3) in the Cartesian coordinates $u$, $x$ of Eq. (4) and considering the limit as $u \to \infty$.

An analogous computation gives similar results for the "past" ($=-T$) directed timelike curves $T = -X + k$ for constant $-\infty < k < \infty$. We recover the second expression of Eq. (3):

$$r = \frac{k}{\sin(\frac{\theta}{2}) + \cos(\frac{\theta}{2})} = \frac{|k|}{\sqrt{1+\sin\theta}} \tag{16}$$

---

‡ The second equality requires the trigonometric half angle identity, $\sin\theta = 2 \sin(\theta/2) \cos(\theta/2)$.

Once again, two lightlike curves of the source Minkowski spacetime, $T = -X + k$ and $T = -X - k$ for $k > 0$ are mapped under Eq. (8) to give a single curve of Eq. (3). These curves are illustrated in Figure 8.

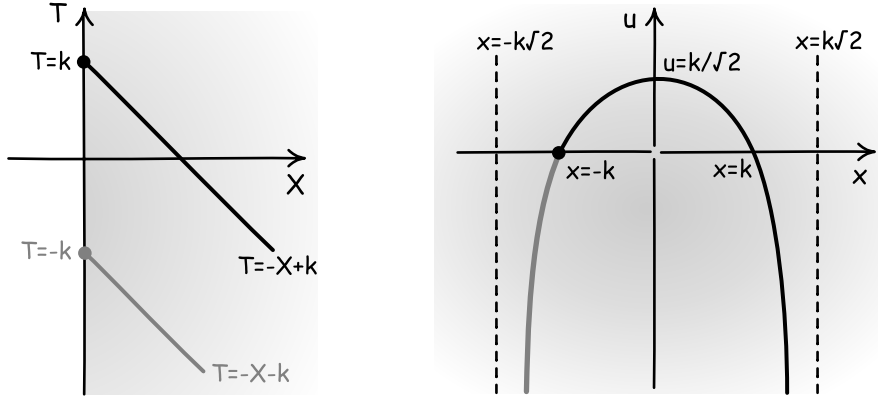

Fig. 8. Second mapping of lightlike curves

The resulting curve in the target spacetime diverges in the $-u$ direction and is also asymptotic to $x = \pm k\sqrt{2}$.

The special case of $k = 0$ corresponds to the two lightlike curves $T = X$ and $T = -X$. The formulae Eq. (3) are degenerate for them. It is easy to see however that these two curves map under Eq. (8) to the lightlike curves $\theta = \pi/2$ and $\theta = -\pi/2$, that is, curves that lie on the $u$ axis.

A check on the consistency of these results for lightlike curves employs the line element (1). If we set $ds^2 = 0$, the line element provides a differential equation that characterizes lightlike curves in the $r$, $\theta$ plane:

$$\left(\frac{dr}{d\theta}\right)^2 - r\tan\theta \frac{dr}{d\theta} - \frac{r^2}{4} = 0 \tag{17}$$

Some manipulations affirms that the expressions Eq. (3) each solve this equation.

## 6. Recovering Timelike Geodesics

This same mapping makes recovery of timelike geodesics straightforward. In the source Minkowski spacetime of Eq. (6), a family of geodesics is defined by $X = k$, that is,

$$r \cos \phi = k \tag{18}$$

for fixed values of $Y$ and $Z$ and for $k > 0$. These correspond under Eq. (8) to

$$r \cos(\theta/2) = k \tag{19}$$

in the target spacetime. Figure 9 shows a geodesic in the source Minkowski spacetime on the left and in the target spacetime on the right.

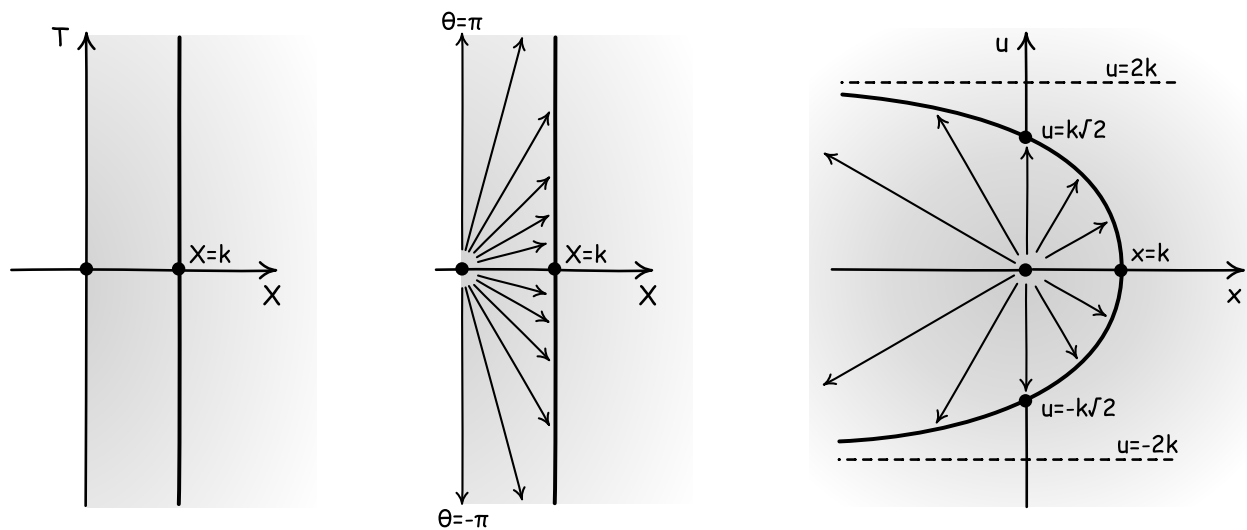

Figure 9. Timelike geodesics

The construction in the center shows how one can recover qualitative features of the transformed geodesic without calculation by inspection of the figure. It shows radial lines of constant $\theta$ that identify a timelike geodesic in the source Minkowski spacetime. The image of the radial lines and the timelike geodesic in the target spacetime is shown on the right of Figure 9. The two radial lines $\theta = \pi$ and $\theta = -\pi$ in the source Minkowski spacetime are parallel to the geodesic. Since these two radial lines coincide with the -x-axis in the target spacetime, we can conclude that the geodesic mapped to the target spacetime will approach lines parallel to the -x-axis for very negative x. Similar interpretations can be applied to Figures 7 and 8.

## 7. The Singularity

The status of the target spacetime at $r = 0$ is, so far, unclear. That the polar coordinates "go bad" at $r = 0$ may merely be an artifact of that coordinate system and may not represent a pathology of spacetime. Such a benign result arises when polar coordinates are used in a Euclidean space. That is not the case here. There is a singularity in the metrical structure of Eq. (1) at $r = 0$. It is the type of singularity that is found in the intrinsic geometry of a cone.

The simplest cone singularity is produced, figuratively, by taking a flat sheet of paper, excising a pie shaped segment and connecting the exposed edges to form a cone, as shown in Figure 10. There is an extrinsic curvature singularity at the pointy apex. The intrinsic geometry of the surface of the cone remains everywhere flat. However, something also goes awry in the intrinsic geometry at the apex. The familiar way to illustrate it, is to note that the circumference of a circle, centered on the apex, no longer obeys the Euclidean result of (circumference) = $2\pi$(radius).

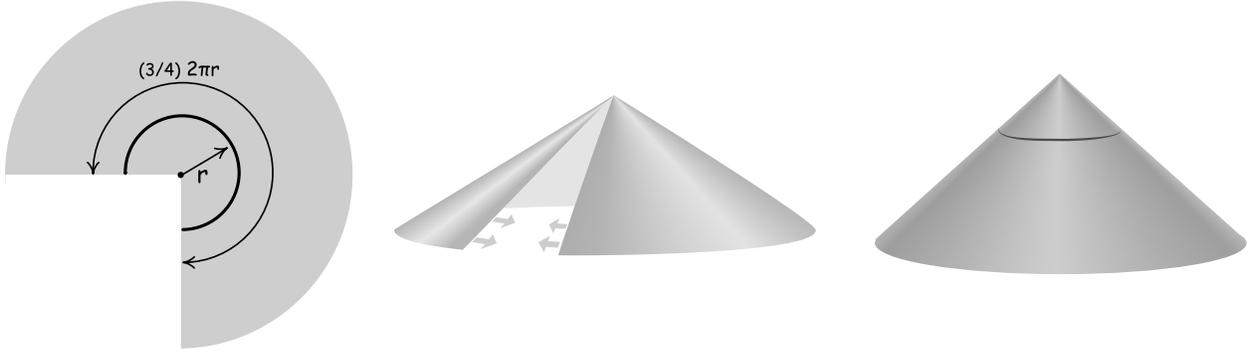

Fig. 10. The simplest cone singularity

In Figure 10, the cone is formed after a quarter of the circle has been excised. If instead a half of the circle had been excised, then we could use the transformation $\theta = 2\phi$ of Eq. (8) to represent the formation of the cone. That is, we use a standard radial coordinate $r$ and angle coordinate $\phi$ on the original sheet of paper, where $r=0$ is the center for the circle. Its Euclidean line element is $ds^2 = dr^2 + r^2 d\phi^2$. Since half the sheet is excised, the remaining half is covered by the angle coordinate $\phi$ in the interval $0 \le \phi \le \pi$. When the exposed edges of the sheet are joined to form the cone, the points with coordinates $\phi = 0$ and $\phi = \pi$ (and equal radial coordinate $r$) are identified. To ensure that an angle coordinate on the surface takes on values in the range 0 to $2\pi$, we replace the old angle coordinate $\phi$ by the new angle coordinate $\theta = 2\phi$ of Eq. (8). The radial coordinate $r$ remains unchanged. Under this transformation, the line element for the cone is

$$ds^2 = dr^2 + r^2 d\phi^2 = dr^2 + r^2 d(\theta/2)^2 = dr^2 + r^2/4 \, d\theta^2$$

This last transformation is analogous to the transformation above from Eq. (10) to Eq. (1).

A fuller analysis shows that the singular character of the geometry *intrinsic to the cone's surface* resides in a failure of uniform convergence of its metrical properties. The pathology is not limited to a curvature singularity in the surface's *extrinsic* curvature.§ That is, if we seek to assign a metric to the apex by taking the limit of the metrical structure on radial paths leading to the apex, we find different metrics according to the radial path chosen.

---

§ This is not assured. A pathology in the extrinsic geometry may not be reflected in the intrinsic geometry. A sharp, linear crease in the sheet of paper is singular in its extrinsic curvature, but remains regular in the intrinsic geometry.

Since the construction of the time-travel spacetime is similar to that of the cone, the same sort of cone singularity arises at $r = 0$ in Eq. (1). Spacetime singularities of this type have been investigated by Ellis and Schmidt[8]. That something is amiss at $r = 0$ follows if we seek to assign light cones to events at $r = 0$. At all regular events in the spacetime where $r > 0$, timelike curves through the event form the familiar double cone, no matter how close that event is to $r = 0$. If we collect the timelike curves converging towards an event at $r = 0$, they form a single cone. Whatever metrical structure we might try to assign to events at $r = 0$, that structure will be unlike the metrical structure at all neighboring events since it must produce a single-lobed lightcone.

The construction of this single-lobed cone is shown in Figure 11. The left of the figure shows timelike curves in the source Minkowski spacetime that will transform to this single cone in the target spacetime. The right of the figure shows these curves after they are transformed under Eq. (8) in the target spacetime.

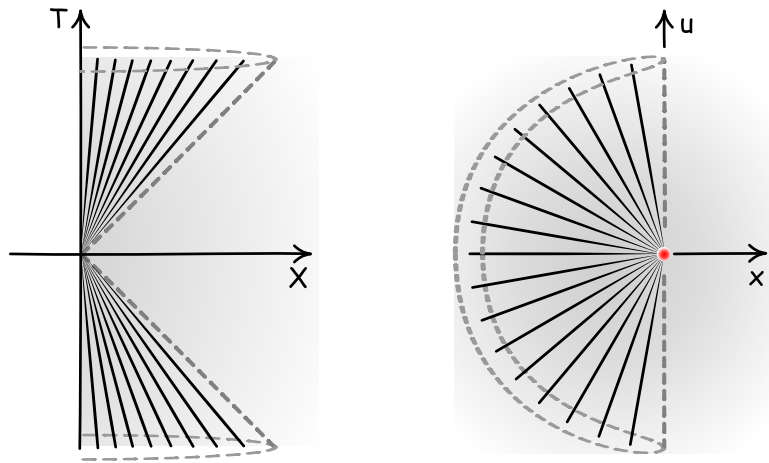

Fig. 11. Degenerate light cone structure

Proceeding more fully, we cannot use the cylindrical coordinate system of Eq. (1) to show the singular character of Eq. (1) at $r = 0$, since that coordinate system is badly behaved at $r = 0$. Instead, we *stipulate* that there are manifold points at $r = 0$. We seek to investigate the metrical structure there using the Cartesian coordinate system $u, r, y, z$, defined in Eq. (4), since that coordinate system is regular at $r = 0$.

We transform the line element Eq. (1) to this new coordinate system using differentials derived from Eq. (4):

$$d\theta = (x/r^2)\, du - (u/r^2)\, dx \qquad dr = (u/r)\, du + (x/r)\, dx \tag{20}$$

After considerable manipulation, the line element Eq. (1) transforms to**

$$ds^2 = \frac{x}{r}\left[\frac{x^2/4}{u^2+x^2}\right]du^2 - \frac{x}{r}\left[\frac{\frac{3u^2}{4}+x^2}{u^2+x^2}\right]dx^2 - \frac{u}{r}\left[\frac{u^2+\frac{3x^2}{2}}{u^2+x^2}\right]dudx - dy^2 - dz^2 \tag{21}$$

where, as before, $r = (u^2 + x^2)^{1/2}$. Using Eq. (4) and with some manipulation, this form of the line element can be rewritten in terms of $\theta$ as

$$ds^2 = \cos\theta\,[(\cos^2\theta)/4]\, du^2 - \cos\theta\,[(3/4)\sin^2\theta + \cos^2\theta]\, dx^2$$
$$- \sin\theta\,[\sin^2\theta + (3/2)\cos^2\theta]\, dudx - dy^2 - dz^2 \tag{22}$$

Its metrical coefficients in a coordinate basis $x^i = (u, x, y, z)$ are

$$g_{ik} = \begin{bmatrix} \cos\theta(\cos^2\theta)/4 & -(\sin\theta)/2\left[\sin^2\theta + (\tfrac{3}{2})\cos^2\theta\right] & 0 & 0 \\ -(\sin\theta)/2\left[\sin^2\theta + (\tfrac{3}{2})\cos^2\theta\right] & -\cos\theta\left[(\tfrac{3}{4})\sin^2\theta + \cos^2\theta\right] & 0 & 0 \\ 0 & 0 & -1 & 0 \\ 0 & 0 & 0 & -1 \end{bmatrix} \tag{23}$$

The distinctive property of this form of the line element is not so much the specific values that these metrical coefficients take. Rather it is just that these coefficients are functions of $\theta$ only and they are different for different values of $\theta$.

This fact reveals the character of the singularity at $r=0$. We may try to define a metrical structure at $r = 0$ in the time-travel spacetime by the requirement of continuity with the metrical structure at neighboring events. That is, we seek to assign a metric to the event $r=0$ as the limit of the metric taken along a constant $\theta$, radial path, terminating in $r=0$. It now follows that this requirement of continuity produces a different metrical structure according to the radial path of constant $\theta$ along which we approach $r = 0$.†† The singular character of the metrical structure at $r = 0$ resides in its necessary discontinuity with the metrical structure of neighboring events.

---

** This formula explains why the lightcones in Figure 2 appear distorted. If we solve for $ds^2 = 0$, we find for light cones on the $x$-axis, where $u=0$, that $du/dx = du/dy = du/dz = \pm 2$. The distortion is a coordinate artifact.

†† These metrics, with different values of $\theta$, are isometric, since all are flat. If realize all of them at the origin, however, they will assign different norms to the same vector. The vector (1, 0, 0, 0) will be assigned the norm $\cos\theta\,[(\cos^2\theta)/4]$, for example. This is an invariant failure to agree.

## 8. Conclusion

What do we learn from this example? In my view, we reaffirm a familiar result: that the spacetimes of general relativity admit time travel. Whether one finds this example more or less illuminating than others is, in the end, decided by what each of us finds more or less natural or intuitive. In this regard, I find it appealing since the spacetime is locally everywhere flat excepting for the singular surface whose presence makes the difference between an everywhere flat Minkowski spacetime without time travel and one with time travel.

One possible reaction—heard anecdotally—is that this spacetime is somehow lesser since it is "unphysical" or, in a different but related concern, "artificial." This notion of being "physical" is an important part of the pragmatics of practical physics. It is invoked to dismiss some particular result from consideration in the particular context at hand.

In another application, four different, precise senses for the notion have been identified.[9] None apply here in so far as the goal is merely to explore the range of spacetimes admitted by general relativity. There is no proposal that this form of time travel is realized in our universe. The situation is analogous to the intrinsic geometry of the cone of Figure 10. It may or may not be the geometry of our space. Nonetheless, its analysis lies within the scope of geometry and that is underscored by the fact that we can build something close to it in a paper model. This example is connected to what is possible in our world by a slender thread: it is a model of our current best theory of space and time, general relativity. In that, however, it keeps company with many other much stranger models.

## Acknowledgments

My thanks to John Earman, John Manchak and David Malament for helpful comments on an earlier draft; and to Bob Wald for helpful discussion on characterizing conical singularities.

The author has no conflicts to disclose.

---

[1] New York: Henry Holt & Co., 1895.

[2] New York: Philomel, 1993.